\documentclass{article}

\usepackage[margin=1in]{geometry}
\usepackage{amsmath,amssymb}
\usepackage{booktabs}
\usepackage{array}
\usepackage{makecell}
\usepackage{tabularx}
\usepackage{enumitem}
\usepackage{graphicx}
\usepackage{hyperref}
\usepackage{xcolor}
\usepackage{needspace}
\usepackage{etoolbox}
\usepackage[font=small,labelfont=bf,skip=6pt]{caption}

\pretocmd{\section}{\Needspace{10\baselineskip}}{}{}
\pretocmd{\subsection}{\Needspace{6\baselineskip}}{}{}
\pretocmd{\paragraph}{\Needspace{4\baselineskip}}{}{}

\hypersetup{
  colorlinks=true,
  linkcolor=blue,
  citecolor=blue,
  urlcolor=blue
}

\newcommand{\modelname}{\textsc{marimba}}

\title{\modelname: A System One Model for Speech}
\author{Solano Amadori\\\nolinkurl{charcot.ai}}
\date{November 2026}

\begin{document}
\maketitle

\begin{abstract}
Speech applications require decisions about term presence, turn completion,
and language within an audio window.
We present \modelname{}, an interface for typed speech
decisions built on frozen Qwen3-ASR~0.6B features. Lexical questions use
zero-shot scoring of ASR transcripts; non-lexical questions use a
question-conditioned head and a separate layer-18 turn-end readout.
Language identification uses the ASR model's own language scores.
% src: CPU audit of saved lexical predictions@7f54a303; maximum binary test ECE=0.07621584915430071 (ID held-out).
On private conversational test sets, decoder AUROC is 0.80--0.92; on
public read speech it is 0.92--0.97. Speech Commands four-way accuracy
is 0.93; binary lexical ECE is at most 0.08.
Personal-information detection, which passes the greedy transcript to a
private, curated PII pipeline, reaches item-level F1 0.904 on synthetic
Brazilian Portuguese speech (1.000 on reference transcripts).
Layer-12 speaker readouts detect speaker change with AUROC 0.966--0.980
on AMI and VoxConverse, above an ECAPA baseline, and count speakers
(0, 1, 2, or 3+) more accurately than pyannote~3.1.
The turn-end readout reaches AUROC 0.930 versus Smart Turn v3.2 at 0.982
on Smart Turn's own test set. On a subset of ICSI meetings labeled by
next speaker, AUROC is 0.716 versus 0.712; the paired confidence
interval includes zero.
\end{abstract}

\noindent Code: \href{https://github.com/Charcot-AI/marimba-api}{\texttt{Charcot-AI/marimba-api}};
weights: \href{https://huggingface.co/charcot-ai/marimba}{\texttt{charcot-ai/marimba}}.

\section{Introduction}

Speech applications require distinct decisions about the same short
window. These include turn-end detection for voice agents, term spotting,
voice activity detection, and language identification. Term spotting tests
whether a queried word was spoken; voice activity detection separates speech
from silence and noise. Constrained answers with calibrated uncertainty can
inform acceptance or deferral for review, but calibration alone does not
validate an automated acceptance policy.

Audio language models, including
Qwen2-Audio, SALMONN, and Qwen3-Omni, accept audio and answer through text
generation without a separate speech-recognition step
\cite{qwen2audio,salmonn,qwen3omni}. They typically require a separate
prompt and generation for each question, and their token log probabilities
are not calibrated probabilities over a typed decision space. Passing a
single 1-best transcript to a text model discards the recognizer's
uncertainty; a rare name or term lost in decoding cannot be recovered or
calibrated downstream. Single-task tools do not provide a shared
request-time interface across terms,
turns, voice activity, and language identification.

Jev, a typed decision model, answers with probabilities over declared
options rather than free text, but does not accept audio \cite{typesafejev}.

\begin{figure}[t]
\centering
\includegraphics[width=\textwidth]{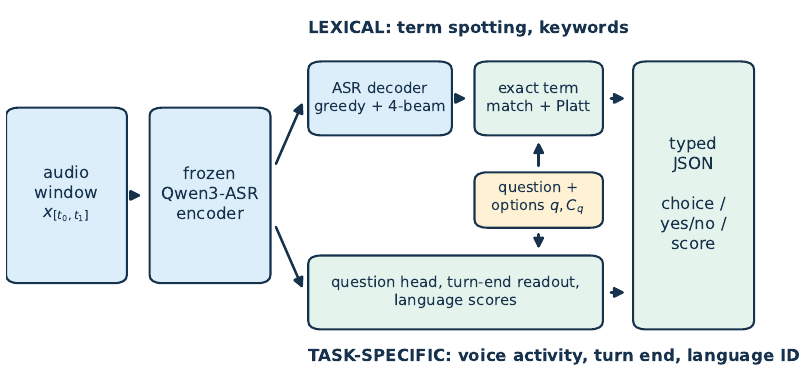}
\caption{The \modelname{} interface separates zero-shot lexical scoring,
question-conditioned non-lexical prediction, a frozen layer-18 turn-end
readout, and the ASR model's language scores. The shared encoder does not
remove the decoding step for lexical questions.}
\label{fig:overview}
\end{figure}

\modelname{}\footnote{The name refers to the percussion instrument; the model
is unrelated to the Mamba state-space architecture.} applies typed decisions
to speech (Figure~\ref{fig:overview}). A frozen Qwen3-ASR encoder processes
each 8-second window once. Lexical questions are scored zero-shot from the
decoder's greedy and four beam-search hypotheses and calibrated with Platt scaling;
non-lexical questions use a question-conditioned head or a frozen
layer-18 turn-end readout. The ASR model's own language scores provide language
identification. We evaluate on public read speech,
meetings, broadcast conversation, and noise, and on private
Brazilian Portuguese medical consultations, which contain rare medication
names, phonetic near-misses, and multi-party turns.

Our contributions are:
\begin{itemize}[leftmargin=*]
  \item A shared interface separating lexical scoring, non-lexical
  prediction, language identification, and a frozen turn-end readout.
  \item A typed task suite covering term spotting, personal-information
  presence, turn end, voice activity, language identification,
  same speaker, speaker change, and speaker count, with
  phonetically confusable hard negatives (including real medication-confusion
  pairs) and terms withheld from training.
  \item A calibration and selective-prediction evaluation reporting AUROC, ECE,
  and accuracy-targeted coverage, with source-specific uncertainty and
  explicit identification of missing measurements.
\end{itemize}

\section{Related work}

\paragraph{Typed decisions.}
Jev produces typed outputs (choices, binary yes/no answers, and scores)
instead of free-form strings. It does not accept audio, according to its
public model documentation \cite{typesafejev}. We do not claim to
reproduce its architecture, training recipe, or RLCD. JevAdvBench
examines benchmark and adversarial robustness, and TypedBench evaluates
calibration, framing sensitivity, and decision cost
\cite{jevadvbench,typedbench}; we likewise evaluate calibration directly.

\paragraph{System One models beyond Jev.}
The closest speech precedent is ASOM. Frozen speech and text encoders feed
a shared reader without per-class parameters. The reader answers request-time
choice, score, and yes/no questions in one pass without generation. Its evaluation is
limited to paralinguistic emotion on MSP-Podcast, including reworded questions
and new label sets \cite{asom}. Our model addresses a broader set of speech decisions on
conversation-disjoint splits, with Portuguese medical consultations as a
stress test for exact lexical checks against held-out medication terms and
real phonetic-confusion pairs. Typed, calibrated speech decisions are not new.

Laya provides open-weight, text-only System One models based on
ModernBERT-large and mmBERT-base and uses RLCD \cite{laya}. Narrow Decision
Models describes an open architecture using block attention for option-order
invariance \cite{narrowdecision}. UniJev studies trimodal text, image, and
video decisions and does not include audio \cite{unijev}.

\paragraph{Frozen speech encoders and task heads.}
Whisper and Qwen3-ASR provide multilingual speech encoders whose features can
be reused by lightweight task heads without autoregressive decoding;
Qwen3-Omni builds on a related audio encoder (AuT). \modelname{} keeps the
Qwen3-ASR~0.6B encoder and decoder frozen \cite{whisper,qwen3asr,qwen3omni}.

\paragraph{Audio language models.}
Qwen2-Audio, SALMONN, and Qwen3-Omni pair audio encoders with language models.
They answer directly from audio without a separate speech-to-text stage
\cite{qwen2audio,salmonn,qwen3omni}, typically generating text from a separate
prompt for each question. Option-token log probabilities can provide candidate
scores, not calibrated typed probabilities by default. We evaluate typed
probabilities separately from discrimination and compare turn-end detection
with a task-specific system.

\paragraph{Single-task speech tools.}
Smart Turn v3.2 is a task-specific turn-end system \cite{smartturn}. Keyword
spotters and voice-activity detectors target word spotting and speech
activity, respectively.

\section{Task formulation}

\subsection{Typed questions}

The \modelname{} interface specifies three output types:
\texttt{choice}, \texttt{yes/no}, and \texttt{score}.

Let \(x\) be an audio segment, \(q\) a question, and \(C_q\) the answer
options permitted by the question. The model receives the audio state and
question together and emits a typed distribution. For a categorical
\texttt{choice} question, the output is
\[
  p_\theta(c \mid x,q), \qquad c \in C_q, \qquad
  \sum_{c \in C_q} p_\theta(c \mid x,q)=1.
\]
For a binary \texttt{yes/no} question, the output is a probability for the
affirmative answer, with its complement assigned to the negative answer. For
a \texttt{score} question, the output is a bounded scalar or a distribution
over a task-defined ordered scale. The machine-readable output contains no
generated rationale. We evaluate and calibrate confidence against held-out
labels; it does not guarantee correctness.

The tasks are term presence, turn end, voice activity, language
identification, same speaker, speaker change, and speaker count.
The output format does not guarantee performance on
every task that can be phrased as a question. Exact medication decisions are
evaluated against phonetically confusable negatives, rather than
semantic similarity.

\subsection{Audio windows}

\begin{figure}[t]
\centering
\includegraphics[width=\textwidth]{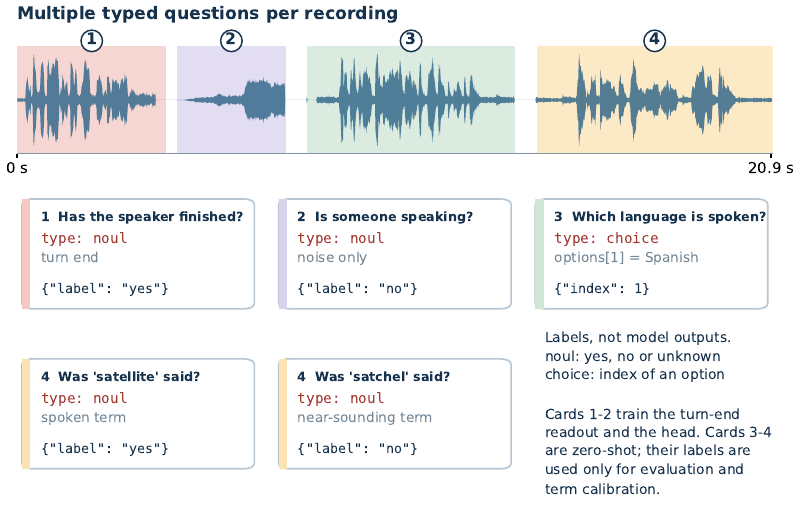}
\caption{Typed questions over one recording built from public clips: FLEURS \texttt{en\_us} and \texttt{es\_419} utterances (CC BY 4.0) \cite{fleurs} and MUSAN noise. Each question selects an interval that is read as one 8\,s window. Each card shows the window's label in the training-data format, not a model output. The turn-end readout and the question head train on such labels; language and term questions are zero-shot, so their labels are used only for evaluation and term calibration.}
\label{fig:windowing}
\end{figure}

Inputs are mono 16\,kHz audio with an eight-second context, but window
construction is task-specific (Figure~\ref{fig:windowing}). Lexical labels
refer only to audio inside the selected window. The turn-end readout retains the
final eight seconds, left-pads shorter inputs, and masks padding during
pooling. Reuse across questions applies to the same encoded window,
not to incremental streaming reuse. Appendix~\ref{app:windows} gives details.

\subsection{Option representation and training}

For a question with textual options, the text encoder represents each option
alongside the question. The audio head scores each candidate conditioned on
the same audio features. During training, option order is shuffled for each
example so that a class is not tied to a fixed position. Training and
evaluation retain option text rather than option indices. The decoder
handles term questions (Section~\ref{sec:decoder}). A fixed list of lexicon
terms is withheld from the head's training data and evaluated separately.

\subsection{Term labels and hard negatives}

Typed templates instantiate content questions from each training transcript.
Labels follow exact token matches. Difficult negatives come from the lexicon
and its confusion pairs, excluding words spoken in the clip. A fixed,
versioned list of terms is withheld from training to evaluate generalization.

Binary examples include positive aligned occurrences and difficult negatives.
The negative pool includes real medication-confusion pairs from a drug
lexicon, corrected-versus-original text pairs where available, and absent
terms; unrelated speech alone would not test whether a model distinguishes
near-sounding medication names. We record label provenance for every example
and report evaluation subsets separately by source and label type.

\section{Data}

\subsection{Public multi-domain data}

Lexical interpretation is zero-shot; training data apply only to the head,
which uses private conversational speech and public data for domain and noise
diversity. Public evaluation
uses Common Voice \cite{commonvoice}, FLEURS \cite{fleurs}, Speech
Commands \cite{speechcommands}, AMI \cite{ami}, VoxConverse
\cite{voxconverse}, and MUSAN from OpenSLR~17 (CC BY 4.0) \cite{musan};
Smart Turn v3.2 supplies turn-end training and test data \cite{smartturn}.
ICSI transfer uses official headset mixes and meeting annotations \cite{icsi}.
FLEURS and Common Voice evaluation includes only clips no longer than 8\,s,
so every transcript describes audio inside the window.

\subsection{Private conversational data}
\label{sec:private-data}

We use private conversational speech, predominantly Brazilian Portuguese,
for research training and evaluation and report only aggregate results.
These data are not released. We check split isolation against each
experiment's immutable data revisions, not a dataset's latest state.

\section{Models}

\subsection{Question-conditioned typed head}

For non-lexical decisions, the frozen Qwen3-ASR encoder produces frame-level features
\(H=f_{\mathrm{audio}}(x)\). A question-and-option encoder represents
\(q\) and each candidate \(c\). A cross-attention scorer computes
\[
  s_c = g_\phi(H, q, c), \qquad
  p(c\mid x,q) = \frac{\exp(s_c/T)}
  {\sum_{c' \in C_q}\exp(s_{c'}/T)} ,
\]
where \(T\) is selected on validation data. The categorical implementation
uses a shared scorer for all candidates, so shuffling candidate order does
not change the mapping from semantic option to output probability. Binary
questions use the corresponding binary distribution. A \texttt{score}
head uses a task-specific output mapping and is evaluated with a proper
scoring rule appropriate to its scale. For multiple questions about one
window, the frozen encoder produces \(H\) once. In a shared inference pass,
the head scores each question with its options and returns one typed
distribution per question.

The voice-activity head was evaluated with Portuguese prompts and with held-out Arabic, Hindi, Korean, and Russian prompts.
Turn-end detection uses the separate readout
below; language identification uses the ASR model's language scores. Speaker decisions use the
readouts in Section~\ref{sec:speaker}. Lexical questions use decoder scoring
(Section~\ref{sec:decoder}).

Training uses the option-shuffling and held-out-term protocol described
above. We split examples by conversation before sampling for each task.
Training uses cross-entropy for categorical choices, binary cross-entropy
for binary decisions, and an appropriate proper loss for score outputs.
We optimize the question-conditioned head while keeping the encoder frozen.
Calibration uses validation data only, never the final test set.

\subsection{Decoder scoring for lexical decisions}
\label{sec:decoder}

Lexical questions use the frozen ASR decoder, without term-specific
training.

For term and keyword questions, the frozen Qwen3-ASR
decoder transcribes the window once with greedy decoding and with
four-beam search, keeping the four best hypotheses. Each hypothesis and the queried term are normalized
(NFKD, accents removed, non-Latin letters kept, lowercase). The raw score is
the fraction of the five hypotheses that contain the normalized term as an
exact token sequence:
\[
  r(x,t) = \frac{1}{|\mathcal{H}(x)|}\sum_{h\in\mathcal{H}(x)}
  \mathbb{1}\bigl[\,\mathrm{norm}(t)\subseteq\mathrm{norm}(h)\,\bigr].
\]
A logistic regression fitted on a held-out calibration split maps \(r\) to a
probability. A \texttt{choice} over terms maximizes the exact-match score,
using fuzzy similarity only to break ties. Exact matching keeps near-miss inflections (e.g., \emph{coloca} versus
\emph{colocar}) as negatives. The
transcription is shared by all lexical questions about the same window, so
adding a term does not add a decoder pass.

Personal-information questions (\texttt{yes/no}) pass the greedy hypothesis
to an external text PII detector and return its binary answer, without a
calibrated probability. The release does not include a detector.

\subsection{Frozen turn-end readout}

% src: docs/turn_end_readout.md at 0daa25886a43c313f14a2ce77b531b9cb4eb52d9, lines 17--58
We fit a logistic readout on layer-18 encoder features, pooling the final
two seconds of valid frames from the same eight-second encoder pass.
Pooling and regularization are selected by five-fold grouped
cross-validation on 4,000 new training clips. The selected tail readout
uses \(C=0.01\) and is refit on 5,768 clips. We fit Platt calibration
on a separate 2,000-clip pool and freeze the configuration and weights
before the 500-clip public test.

\subsection{Speaker readouts}
\label{sec:speaker}

% src: docs/diarization_preregistration.md; artifacts/diarization/selection.json@924cb9ef
Same-speaker, speaker-change, and speaker-count decisions use small
readouts on layer-12 encoder features from the same eight-second encoder
pass. Same speaker compares the two four-second halves, each with one
active speaker; speaker change compares the dominant speaker in the last
two seconds with the preceding six; speaker count predicts 0, 1, 2, or 3+
speakers with at least 0.5\,s of speech. Same speaker and speaker count use
logistic regression on pooled frames; speaker change uses a one-hidden-layer
MLP. Labels come from human speaker timestamps in AMI and VoxConverse,
split by recording. Layer, readout, and Platt or temperature calibration
are selected on validation only and frozen before the test.

\subsection{Language identification}

% src: docs/language_id_native.md, lines 41--72
We sum conditional log probabilities of Qwen3-ASR's built-in language-name
tokens, without an adapter or separate language head. Open-set prediction
considers the 30 supported languages; closed-set prediction restricts
scores to the question's options. Confidence is the softmax over the options.

\section{Experiments}

We report lexical scoring (Table~\ref{tab:decoder-final}), turn-end
detection (Table~\ref{tab:turn-end}), and language identification
separately. Table~\ref{tab:head-final} reports the remaining tasks.

\label{sec:results}

\begin{table}[t]
\centering
\small
\caption{Final decoder results for zero-shot Qwen3-ASR~0.6B (no
fine-tuning). ID: held-out conversations from the training corpus release;
Out-of-domain and Curated: private consultation corpora not used for training;
Gold: reviewed subset of Curated. Seen and held-out refer to the head's term
list; the decoder is trained on neither. Logistic (Platt) calibration is fitted on
a held-out calibration split; 95\% confidence intervals use bootstrap resampling clustered
by audio clip. Common Voice and FLEURS use a versioned public evaluation release,
filtered to clips of at most 8\,s. Four-way accuracy is followed by the number
of choice questions.}
\label{tab:decoder-final}
\begin{tabular}{l l r c c c r}
\toprule
Test & Terms & \(n\) & AUROC [95\% CI] & ECE & Cov@95 & 4-way acc. \\
\midrule
% src: saved ASR qwen3_asr_0_6b@7f54a303 (prep 9ab1b258); clip bootstrap 1000/seed=0; ID held-out.
ID & held-out & 71 & 0.885 [0.817, 0.946] & 0.08 & 0.63 & 0.85 (39) \\
% src: saved ASR qwen3_asr_0_6b@7f54a303 (prep 9ab1b258); clip bootstrap 1000/seed=0; ID seen.
 & seen & 797 & 0.876 [0.845, 0.907] & 0.03 & 0.74 & 0.88 (178) \\
\addlinespace
% src: saved ASR qwen3_asr_0_6b@7f54a303 (prep 9ab1b258); clip bootstrap 1000/seed=0; Out-of-domain held-out.
Out-of-domain & held-out & 2,426 & 0.876 [0.862, 0.889] & 0.06 & 0.38 & 0.83 (1,104) \\
% src: saved ASR qwen3_asr_0_6b@7f54a303 (prep 9ab1b258); clip bootstrap 1000/seed=0; Out-of-domain seen.
 & seen & 23,981 & 0.861 [0.853, 0.868] & 0.03 & 0.27 & 0.81 (5,439) \\
\addlinespace
% src: saved ASR qwen3_asr_0_6b@7f54a303 (prep 9ab1b258); clip bootstrap 1000/seed=0; Curated held-out.
Curated & held-out & 711 & 0.899 [0.876, 0.921] & 0.07 & 0.52 & 0.88 (397) \\
% src: saved ASR qwen3_asr_0_6b@7f54a303 (prep 9ab1b258); clip bootstrap 1000/seed=0; Curated seen.
 & seen & 5,633 & 0.896 [0.881, 0.909] & 0.05 & 0.97 & 0.86 (1,186) \\
\addlinespace
% src: saved ASR qwen3_asr_0_6b@7f54a303 (prep 9ab1b258); clip bootstrap 1000/seed=0; Gold held-out.
Gold & held-out & 133 & 0.915 [0.865, 0.962] & 0.04 & 0.57 & 0.92 (50) \\
% src: saved ASR qwen3_asr_0_6b@7f54a303 (prep 9ab1b258); clip bootstrap 1000/seed=0; Gold seen.
 & seen & 1,281 & 0.801 [0.770, 0.833] & 0.02 & 0.14 & 0.82 (299) \\
\midrule
% src: saved ASR qwen3_asr_0_6b@7f54a303 (prep 9ab1b258); clip bootstrap 1000/seed=0; Common Voice seen.
Common Voice & seen & 912 & 0.923 [0.905, 0.939] & 0.03 & 0.89 & 0.97 (456) \\
% src: saved ASR qwen3_asr_0_6b@7f54a303 (prep 9ab1b258); clip bootstrap 1000/seed=0; FLEURS seen.
FLEURS & seen & 957 & 0.970 [0.958, 0.979] & 0.03 & 1.00 & 1.00 (477) \\
% src: saved ASR qwen3_asr_0_6b@7f54a303 (prep 9ab1b258); clip bootstrap 1000/seed=0; Speech Commands.
Speech Commands & -- & 2,900 & 0.932 [0.922, 0.941] & 0.01 & 0.73 & 0.93 (1,450) \\
\bottomrule
\end{tabular}

\vspace{4pt}
\begin{minipage}{0.92\textwidth}\footnotesize
% src: saved ASR qwen3_asr_0_6b@7f54a303 (prep 9ab1b258); clip bootstrap 1000/seed=0; Common Voice held-out.
Common Voice held-out (\(n=13\)): AUROC 0.917 [0.750, 1.000], ECE 0.02,
Cov@95 0.92, four-way accuracy 0.83 (6). FLEURS held-out (\(n=3\)): AUROC
undefined.
\end{minipage}
\end{table}

\begin{table}[t]
\centering
\small
\caption{Other release evaluations.}
\label{tab:head-final}
\begin{tabular}{l l c c}
\toprule
Task & Test & AUROC / accuracy & ECE \\
\midrule
Voice activity & AMI / VoxConverse & AMI 0.986--0.995; Vox \(\geq\)0.999 & 0.003--0.028 (cal.) \\% src: release VAD report
Unseen-task scene classification & CochlScene (subset) & 0.48--0.54 / 0.24--0.29 & 0.12--0.49 (raw) \\% src: docs/unseen_task_eval.md, CochlScene table, range over evaluated heads and seeds
\bottomrule
\end{tabular}
\end{table}

\begin{table}[t]
\centering
\small
\setlength{\tabcolsep}{4pt}
% src: docs/turn_end_readout.md at 0daa25886a43c313f14a2ce77b531b9cb4eb52d9, Smart Turn and ICSI tables
\caption{Turn-end detection on matched inputs. AUROC intervals use bootstrap
resampling clustered by clip for Smart Turn (recording IDs unavailable) and meeting
clusters for ICSI. Readout ECE is shown before \(\rightarrow\) after frozen Platt calibration;
Smart Turn v3.2 is not recalibrated.}
\label{tab:turn-end}
\begin{tabular}{l r l c c}
\toprule
Test & \(n\) & Model & AUROC [95\% CI] & ECE \\
\midrule
Smart Turn & 500 & Layer-18 readout & 0.930 [0.906, 0.951] & 0.034 \(\rightarrow\) 0.041 \\
 & 500 & Smart Turn v3.2 & 0.982 [0.973, 0.990] & 0.030 \\
\addlinespace
ICSI & 1,728 & Layer-18 readout & 0.716 [0.669, 0.756] & 0.207 \(\rightarrow\) 0.200 \\
 & 1,728 & Smart Turn v3.2 & 0.712 [0.649, 0.770] & 0.143 \\
\bottomrule
\end{tabular}
\end{table}

\paragraph{Turn-end detection.}
The paired AUROC gap is \(-0.052\), with 95\% CI
\([-0.075,-0.032]\), on Smart Turn's in-distribution test.
Frozen Platt calibration increases
readout ECE from 0.034 to 0.041. The ICSI test set contains
1,728 windows from 27 of 75 meetings, with exclusions due only to
download failures. Labels describe the next speaker, not semantic
completeness. The paired difference is
\(+0.004\), with 95\% CI \([-0.046,+0.054]\): a statistical tie, not evidence of
advantage or equivalence.

\paragraph{Recalibration.}
% src: docs/turn_end_readout.md at 0daa25886a43c313f14a2ce77b531b9cb4eb52d9, cross-meeting recalibration
Five-fold cross-meeting recalibration on the same ICSI windows reduces
readout ECE from 0.200 to 0.011 with 200 labeled windows per fold. This
in-domain analysis does not establish cross-domain calibration.

% src: docs/audio_llm_paired.md, "Release pipeline rerun"
% src: docs/audio_llm_paired.md, release summary.json
% src: docs/pii_eval.md (synthetic pt-BR PII eval; item bootstrap 1000/seed=42)
\paragraph{Personal information.} We synthesized 200 Brazilian Portuguese
clips of at most 8\,s with Piper TTS in four voices. Of these, 100 contain one
name, CPF (Brazilian taxpayer ID) number, phone number, email, or street address in a conversational
sentence, and 100 contain no personal information, including dose, time, and
measurement phrases. With a private, curated PII pipeline applied to
the greedy transcript, item-level precision is 0.966, recall 0.850, and F1
0.904 (95\% CI 0.856--0.945); 3 of 100 negatives are flagged. The same pipeline
scores F1 1.000 on reference transcripts, so the misses come from
transcription: recall is 0.95 for names, 1.00 for emails and addresses,
0.70 for phone numbers, and 0.60 for CPF numbers, whose spoken digits are
often mistranscribed. Run end to end on the audio with its own production
transcriber, the same pipeline scores F1 0.975 (95\% CI 0.949--0.995), with
1 of 100 negatives flagged.

\subsection{Comparison with audio LLMs}
\begin{figure}[tbp]
\centering
\includegraphics[width=\linewidth]{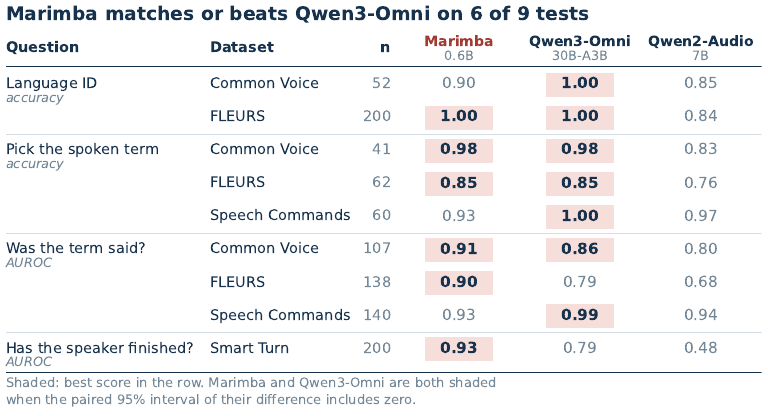}
\caption{Marimba and two zero-shot audio LLMs on the same 1,000 public questions. Language and term choice use accuracy; binary questions use AUROC. Shading follows paired 95\% bootstrap intervals clustered by audio (1,000 replicates): when the interval of the Marimba minus Qwen3-Omni difference includes zero, both are shaded. The turn-end readout was trained on Smart Turn training data; all other rows are zero-shot for every system.}
\label{fig:paired}
\end{figure}

We compare \modelname{} with Qwen3-Omni-30B-A3B-Instruct and Qwen2-Audio-7B-Instruct on 1,000 fixed public questions (Figure~\ref{fig:paired}). Both LLMs score each option zero-shot by summed log-probability. \modelname{}'s lexical and language answers are also zero-shot; its turn-end readout is trained on Smart Turn training data, so that row is in-distribution for \modelname{} only. \modelname{} exceeds both LLMs on FLEURS binary term detection (AUROC 0.901 vs.\ 0.791 and 0.682) and on turn end (200-clip subset; 0.935 vs.\ 0.786 and 0.476). Qwen2-Audio is at chance on turn end (95\% CI 0.40--0.55): it prefers the affirmative option on all 200 clips. Qwen3-Omni is better on Speech Commands (binary AUROC 0.991 vs.\ 0.926; choice accuracy 1.000 vs.\ 0.933) and on Common Voice language identification (accuracy 1.000 vs.\ 0.904). The remaining differences from Qwen3-Omni have 95\% intervals that include zero.

\paragraph{Language identification.}
% src: docs/language_id_native.md, n=300 per corpus, Wilson intervals
On 300 FLEURS examples, open- and closed-set accuracy are both 100\%
[95\% Wilson CI 98.74\%, 100\%]. On 300 Common Voice examples,
open-set accuracy is 84.33\% [79.79\%, 88.01\%] and closed-set accuracy
is 93.33\% [89.93\%, 95.64\%]. ECE is below 0.001
on FLEURS and 0.048 on Common Voice.

\paragraph{Term detection.}
On each private test, the held-out-term AUROC point estimate is at least
as high as the seen-term estimate. This does not establish equal difficulty.
Coverage varies substantially by corpus and summarizes a test-set
confidence curve, not a policy with an independently fixed threshold.
Coverage of 0.97 on curated seen terms is sensitive to test-set composition
and ties. The reviewed seen-term result remains limited by ASR misses
on drug names.

\paragraph{Voice activity.}
Voice activity reaches AUROC 0.986--0.995 on AMI (\(n=474\)) and at least 0.999 on VoxConverse (\(n=4{,}636\)) across three seeds with Portuguese prompts, with calibrated ECE at most 0.028; with held-out Arabic, Hindi, Korean, and Russian prompts, AUROC is at least 0.98.
Speech/music/noise classification is not a released output.

\paragraph{Speaker decisions.}
% src: docs/diarization_results.md@f631d23b; metrics.json (recording-clustered bootstrap)
Table~\ref{tab:speaker} reports the speaker readouts on AMI and VoxConverse
test recordings, scored once. Each readout meets the preregistered criterion of scoring at most 0.05 below its
baseline, with calibrated ECE below 0.035.
Speaker change exceeds ECAPA~\cite{desplanques2020ecapa} on both sets and
pyannote~3.1~\cite{plaquet2023powerset} (0.865 and 0.812). Same speaker
is on par with ECAPA; the AMI result rests on 32
different-speaker pairs. Speaker count exceeds pyannote~3.1 in accuracy
but undercounts three or more speakers: on VoxConverse it predicts two for
395 of 667 such windows.

\begin{table}[t]
\centering
\small
\caption{Speaker decisions on public test recordings (8\,s windows). AUROC
for same speaker and speaker change, accuracy for speaker count, with 95\%
recording-clustered bootstrap intervals. ECE is calibrated (top-label for
count).}
\label{tab:speaker}
\begin{tabular}{llrllr}
\toprule
Task & Test & \(n\) & \modelname{} & Baseline & ECE \\
\midrule
Same speaker & AMI & 380 & 0.998 [0.993, 1.000] & 0.985 ECAPA & 0.020 \\
 & VoxConverse & 7,265 & 0.996 [0.994, 0.998] & 0.995 ECAPA & 0.004 \\
Speaker change & AMI & 751 & 0.966 [0.952, 0.978] & 0.924 ECAPA & 0.034 \\
 & VoxConverse & 10,006 & 0.980 [0.976, 0.984] & 0.962 ECAPA & 0.014 \\
Speaker count & AMI & 960 & 0.778 [0.749, 0.807] & 0.738 pyannote & 0.019 \\
 & VoxConverse & 10,644 & 0.859 [0.844, 0.873] & 0.836 pyannote & 0.007 \\
\bottomrule
\end{tabular}
\end{table}

\paragraph{Unseen tasks.}
CochlScene scene classification is the principal unseen-task evaluation. On a seeded, recording-grouped subset (150 clips per class; not the official test split), every evaluated head is at chance: four-way choice accuracy 0.24--0.29 (\(n=492\), chance 0.25) and binary AUROC 0.48--0.54 (\(n=963\)), with all 95\% intervals including chance.

\paragraph{Latency.}
% src: marimba-api docs/benchmarks.md, End-to-end per-path latency, L4 bf16
On one NVIDIA L4 (bf16, batch size 1, 8\,s windows), end-to-end p95 latency per question, measured around the full public API call, is 13\,ms for turn end, 16\,ms for voice activity, 161\,ms for language identification, and 2.35\,s for term detection with greedy plus four-beam decoding. Lexical scoring requires autoregressive decoding; encoder reuse does not establish streaming readiness.

\begin{figure}[tbp]
\centering
\includegraphics[width=\textwidth,height=0.68\textheight,keepaspectratio]{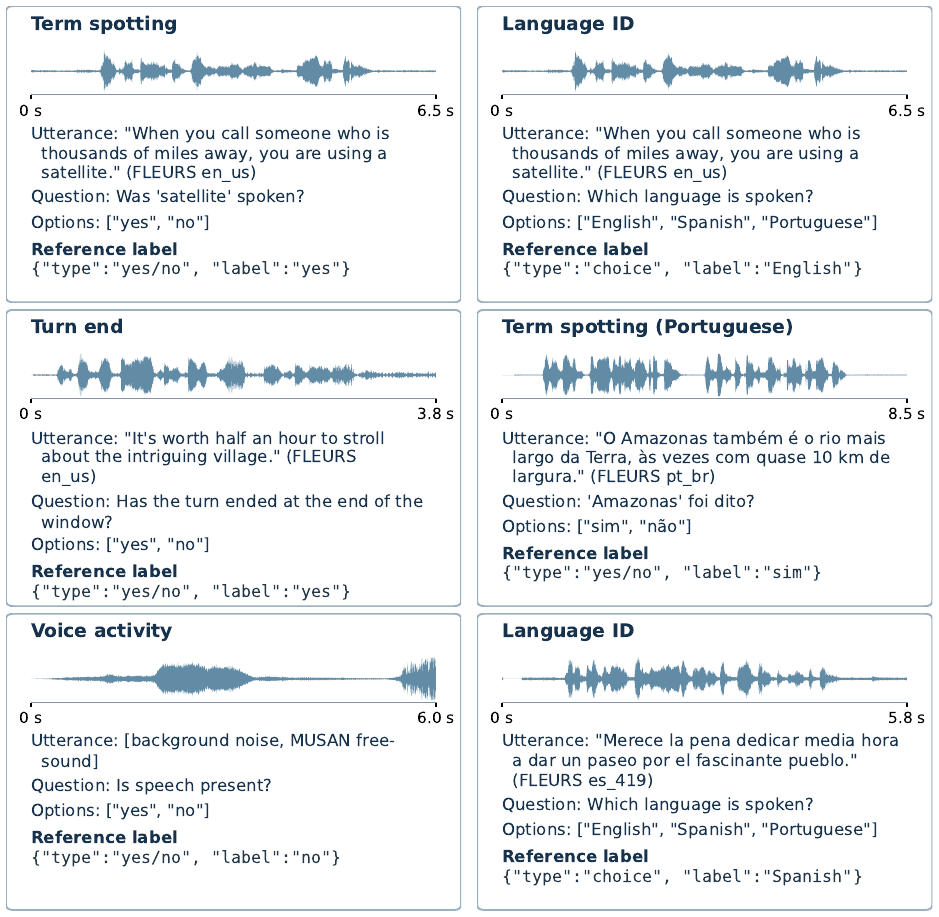}
\caption{Typed questions on public audio: term presence in English and
Portuguese, turn end, voice activity, and language
identification.
Speech comes from FLEURS (CC BY 4.0) \cite{fleurs} and noise from MUSAN
(OpenSLR~17; the displayed noise clip is public domain) \cite{musan}.
Panels show waveforms and dataset-derived reference answers, not model
predictions. The noise example is a negative speech-presence label,
not a sound-classification result.}
\label{fig:decision-panels}
\end{figure}

\subsection{Calibration protocol}

We fit Platt/logistic calibration for the decoder on a held-out calibration
split. For the head, we fit temperature scaling on one calibration subset
and compare it with isotonic regression for binary tasks on a held-out
calibration subset, selecting the method by ECE. Test labels determine
neither calibration parameters nor bin boundaries. Coverage curves describe
test-set results, not independently validated abstention policies.

We report expected calibration error (ECE), stratified by source,
seen versus held-out terms, and decision type. We specify bin definitions
and subgroup criteria before test evaluation, not from test outcomes.

\subsection{Decision thresholds and abstention}

The consuming application chooses a probability threshold according to the
costs of false positives and false negatives. We do not set an automated
action threshold. Deployment evaluation must specify a decision cost matrix,
an escalation path, and a human review policy. Calibration can support this
policy but does not establish deployment suitability.

\section{Limitations and ethics}

Exact term scoring inherits ASR omissions and confusions. Private lexical
results are specific to the evaluated test sets; public read speech does
not establish clinical validity. Speaker decisions are per-window, not full diarization, are measured only on
AMI and VoxConverse, and undercount three or more speakers; voice-activity prompts cover five languages but not English, and CochlScene scene classification is at chance. Personal-information detection is evaluated only on synthetic, templated Portuguese speech and depends on an external text detector. The ICSI test covers
only the 27 meetings we could download, and its labels mark the next speaker, not semantic completeness.
Low in-domain ECE does not establish calibration after deployment shift.

The private recordings were collected through a commercial clinical
documentation platform whose terms of use, accepted by clinicians (the data
controllers under Brazil's LGPD), authorize the provider to de-identify
platform data and use it to train AI models. Consultation and clinician
identifiers are salted before use; recordings are kept in access-controlled
private storage, are not released, and only aggregate results are reported.
The study was not reviewed by a research ethics committee, and patients did
not give study-specific research consent. Public data retain their original
licenses and usage conditions. These results do not validate autonomous
clinical decisions; deployment requires appropriate privacy, legal, and
human-oversight review.

\section{Conclusion}

\modelname{} answers typed questions about a speech window from a single pass of a frozen Qwen3-ASR~0.6B encoder, routing lexical questions to scoring of ASR transcripts and non-lexical questions to a question-conditioned head or a layer-18 readout. Three results support this design. First, lexical scoring requires no term-specific training: on private conversational tests, AUROC is 0.80--0.92 for both seen and held-out terms, and binary ECE is at most 0.08. Second, on 1,000 fixed public questions the interface exceeds Qwen3-Omni and Qwen2-Audio on binary FLEURS term detection, while Qwen3-Omni leads on Speech Commands and Common Voice language identification. Third, a frozen readout approaches a dedicated turn-end model, with AUROC 0.930 versus 0.982 on Smart Turn's own test set and a statistical tie on ICSI. Voice activity reaches AUROC 0.986--0.995 on AMI and at least 0.999 on VoxConverse, and at least 0.98 with four held-out prompt languages. Layer-12 speaker readouts detect speaker change with AUROC 0.966--0.980, above ECAPA, and count speakers more accurately than pyannote~3.1. Passing the greedy transcript to a text PII model detects personal information in synthetic Portuguese speech with F1 0.904.

The same evaluation marks the limits of the approach. Lexical decisions inherit ASR misses on rare terms such as medication names, and every evaluated head is at chance on CochlScene scene classification. Term detection is also the slowest path, at 2.35\,s p95 per question on one L4, against 13--16\,ms for turn end and voice activity.

\bibliographystyle{plain}
\bibliography{refs}

\appendix
\section{Typed interface and evaluation details}

\subsection{Output schema}

An implementation returns the task name, type tag, option set when
applicable, probability distribution or scalar score, and model/version
identifier. The type tag in the output may differ from
the names used here. No natural-language explanation is required for a valid result.
For \texttt{choice}, probabilities are normalized across the currently
presented semantic options. For \texttt{yes/no}, the positive and negative
probabilities sum to one. For \texttt{score}, the scale and calibration
target are task-defined and must be included in the schema. Invalid,
unsupported, or out-of-distribution questions should produce an explicit
abstention rather than a fabricated answer.

\subsection{Window construction}
\label{app:windows}

Audio is converted to mono and resampled to 16\,kHz before any cut.
Lexical intervals longer than 8\,s keep their first 8\,s. Recordings with word
timestamps yield one window at the start and one at a random offset, and a
term counts as spoken only when the whole word lies inside the window.
Recordings longer than 8\,s without word timestamps are skipped rather than
cut, so a term label always refers to audio inside the window.
For the turn-end readout, retain the final eight seconds and left-pad
shorter audio; padding is masked before pooling the last two valid seconds.

\subsection{Training and evaluation separation}

The train/test split is fixed before training examples are sampled. All
windows from a conversation share a split assignment. A separate term list
identifies lexicon terms eligible for training. Validation
serves only model selection and calibration. The turn-end readout uses
separate selection and calibration pools; its public test is not used for fitting.
Cross-meeting recalibration uses out-of-fold predictions and is reported
separately from frozen transfer.

\subsection{Metrics}

For binary questions, AUROC and F1 summarize ranking and a fixed-threshold
operating point. AUROC is computed from pre-calibration scores, because a
step-function calibrator such as isotonic regression can create ties that
change AUROC; accuracy, calibrated ECE, and coverage use calibrated
probabilities. For categorical questions, accuracy and macro-F1 summarize
classification; proper log loss and ECE assess probability quality. For
score outputs, the selected proper scoring rule and calibration diagnostic
must match the declared scale. Results are reported with source-stratified
breakdowns and uncertainty estimates specified before the test run.

Lexical AUROC intervals use 1,000 clip-clustered bootstrap replicates
with seed 0. Lexical ECE uses 15 equal-width bins; turn-end ECE uses 10.
Selective coverage is the
largest fraction of test predictions attaining the target accuracy on
the confidence-sorted curve; ties make it sensitive to test-set composition.

\subsection{Alternative lexical adaptation}

The lexical results use the calibrated zero-shot decoder, not LoRA
adaptation, which fails the calibration and latency criteria.

\end{document}